# From classical to quantum spintronics : Theory of coherent spin injection and spin valve phenomena

Francisco Mireles* and George Kirczenow

*Department of Physics, Simon Fraser University, Burnaby, BC, Canada, V5A 1S6.*

We present a theory of coherent quantum transport in ferromagnetic/ non-magnetic/ ferromagnetic heterojunctions. We predict quantum coherence to give rise to a *quantum* spin valve effect that, unlike its familiar classical analog, occurs even in the absence of a net spin current through the heterostructure. Thus the relationship between spin and charge transport is *qualitatively* different in the presence of quantum interference than in the (semi)classical regime. This has important implications for the design of quantum coherent spintronic devices and the interpretation of experiments.

The physics of injection of spin-polarized electric currents from magnetic materials into semiconductors through solid state interfaces is currently attracting much attention because of its fundamental interest and its potential relevance to the realization of spintronic semiconductor devices that utilize the electron's spin degree of freedom as well as its charge to store, process and transmit information.[1-3] High spin-injection efficiencies have been reported from magnetic to non-magnetic semiconductors at low temperatures.[4,5] Efficient injection of spin-polarized electrons from ferromagnetic metals (F) to semiconductors (S), although desirable for room temperature semiconductor spintronic devices, has not yet been achieved experimentally because the large mismatch between the resistivities of metals and semiconductors is an obstacle to spin injection[6-9], and the interpretation of relevant spin-transport measurements has been controversial.[10-16] However theoretical work has suggested that enhanced spin injection efficiencies may be achieved by fabricating F/S interfaces with suitable energy barriers[7,17] and/or symmetry properties,[18] and experiments on spin injection from ferromagnetic metals to semiconductors via STM and Schottky tunnel barriers have yielded encouraging results.[19-21]

It was suggested in the seminal work of Datta and Das[3] that spin injection may have particularly interesting ramifications in F/S/F double heterojunctions and that a unique transistor that relies on manipulation of the electron's spin instead of its charge may be feasible. However only a few experimental attempts to study spin-transport through F/S/F double heterojunctions have been reported.[10,15,22–24] Calculations of the spin-conductances of two-terminal F/S/F systems have been reported in the semi-classical diffusive and ballistic regimes.[6,25,26] However no theory of transport in F/S/F systems has treated the effects of spin injection and quantum coherence together within the same theoretical framework[27].

In this letter we explore the interplay between spin injection and quantum coherence in ballistic F/S/F systems theoretically within the Landauer formalism of transport.[28] Spin-dependent electron transmission at the interfaces, Rashba spin-orbit coupling[29-32] and quantum interference are treated in a unified way. We find that quantum coherence can have unexpected implications for spin injection and that some intuitive concepts that have played a key role in the development of the field of spintronics but are founded on semi-classical physics no longer apply. For example, we show that in the ballistic quantum coherent regime a pronounced spin valve effect (a change in resistance when the magnetization of a ferromagnetic electrode is reversed) can occur *without any spin-polarization of the current flowing through the semiconductor*. This surprising phenomenon is an inherently *quantum* spin valve effect since it has no analog in the semi-classical ballistic and diffusive transport regimes that have been considered previously. It will be essential to take account of the phenomena that we introduce here in interpreting spin injection experiments in the quantum regime and in schemes for quantum computation that involve spin injection. While we focus here on F/S/F heterojunctions, the quantum spin valve effect that we introduce is general and should also occur in all-metal and all-semiconductor coherent quantum systems.

Let us start by reviewing briefly the definition of the spin-injection rate at a single F/S interface, and of the magnetoconductance (or magnetoresistence) of a F/S/F heterojunction, and how these relate to spin-injection in three-layer systems in the semi-classical regime. The spin-injection rate through a single heterojunction is related to the electric current through the ratio[7,25]

$$\frac{j_M}{j_e} = \frac{j_\uparrow - j_\downarrow}{j_\uparrow + j_\downarrow}\frac{\mu_B}{e} \equiv \eta_M \frac{\mu_B}{e} \qquad (1)$$

where $j_M$ and $j_e$ are the net magnetization current and electric current, respectively, passing through the interface at an applied voltage $V$. $\mu_B$ is the Bohr magneton. The interfacial transport parameter $\eta_M$ describes the degree of spin-polarization of the net electron flux through the interface. The ratio (1), was originally introduced for ferromagnetic metal/paramagnetic metal interfaces, but applies equally to F/S interfaces, assuming that no spin-flip scattering at the interface or spin-precession is present.[25] In terms of the spin-conductances $G_\sigma$, $j_e = j_\uparrow + j_\downarrow = (G_\uparrow + G_\downarrow)V$. Hence $\eta_M = (G_\uparrow - G_\downarrow)/(G_\uparrow + G_\downarrow)$.

For the double interface F/S/F heterojunction we can, with the same assumptions, define the spin-injection efficiency for the whole structure in an entirely analogous way. In terms of the total spin-conductances (assuming that the ferromagnetic contacts have parallel magnetization) it is given by $\eta'_M = (G^{tot}_\uparrow - G^{tot}_\downarrow)/(G^{tot}_\uparrow + G^{tot}_\downarrow)$. In the semi-classical regime where all quantum phase information is assumed to be lost during (ballistic) electron transit between between interfaces, elastic multiple scattering at the interfaces results in spin-transmission probabilities $T^{tot}_\sigma = T_{P\sigma} = T_\sigma/(2 - T_\sigma)$,[26] where $T_\sigma$ are the single-interface transmission probabilities, $\sigma = (\uparrow, \downarrow)$ and $P$ denotes the parallel configuration of the ferromagnets. Hence, semi-classically,

$$\eta'_M = (T_\uparrow - T_\downarrow)/(T_\uparrow + T_\downarrow - T_\uparrow T_\downarrow) \quad (2)$$

and a net spin current flows when $T_\uparrow \neq T_\downarrow$.

A standard although indirect way to detect spin-injection experimentally is based on the spin-valve effect, the change in conductance (or resistance) when the magnetizations of the ferromagnetic contacts switch between the parallel (P) and anti-parallel (AP) configurations. This is represented by the ratio[15,22]

$$\frac{\Delta G}{2G_{av}} = \frac{\Delta R}{2R_{av}} = \frac{G^P - G^{AP}}{G^P + G^{AP}} \equiv \eta. \quad (3)$$

Since the transmission probability for the anti-parallel configuration is $T_{AP} = T_{AP\uparrow} + T_{AP\downarrow}$, and semi-classically $T_{AP\uparrow} = T_{AP\downarrow} = T_\uparrow T_\downarrow/(T_\uparrow + T_\downarrow - T_\uparrow T_\downarrow)$,[26] it follows that

$$\eta = \frac{(T_\uparrow - T_\downarrow)^2}{T_\uparrow^2 + 6T_\uparrow T_\downarrow - 4T_\uparrow^2 T_\downarrow - 4T_\uparrow T_\downarrow^2 + 2T_\uparrow^2 T_\downarrow^2 + T_\downarrow^2} \quad (4)$$

Therefore, as for $\eta'_M$, $\eta$ is not zero when $T_\uparrow \neq T_\downarrow$. Thus (in geometries that exclude extrinsic signals due to local Hall fields and the like[15]) the observation of a spin-valve effect ($\eta \neq 0$) in the semi-classical ballistic regime implies that spin-injection is taking place and vice versa. Similarly, it is generally believed (with the same caveat[15]) that observation of a spin valve effect in the semi-classical diffusive regime indicates that spin injection is taking place and that the same is true for all-metal systems.

We now turn to the coherent quantum regime. We consider spin-transport along the $x$-axis in a quasi-one dimensional wave guide that contains the F/S/F heterojunction and assume that an asymmetric quantum well confines electrons in the $y$-direction in the semiconductor giving rise to a Rashba spin-orbit coupling.[29–32] In the (identical) ferromagnetic electrodes a Stoner-like model of the magnetization is used with the spin-up and spin-down band energies in the electrodes offset by an exchange splitting $\Delta$ (Fig. 1a). The effective mass Hamiltonian for parallel (P) magnetization of the ferromagnets in the $z$ direction is

$$\hat{H} = \frac{1}{2}\hat{p}_x \frac{1}{m^*(x)} \hat{p}_x + \frac{1}{2\hbar}\sigma_z[\hat{p}_x \alpha_R(x) + \alpha_R(x)\hat{p}_x]$$
$$+ \frac{1}{2}\Delta\sigma_z + (\delta E_c - \frac{1}{2}\Delta\sigma_z)\theta(x)\theta(l_s - x) \quad (5)$$

Here $\theta(x)$ is the Heaviside function, and the F/S and S/F interfaces are at $x = 0$ and $x = l_s$, respectively, see Fig. 1(a). The position-dependent electron effective mass is $m^*(x) = m^*_f + (m^*_s - m^*_f)\theta(x)\theta(l_s - x)$, where $f$ and $s$ indicate the ferromagnet and semiconductor regions. We use the one-dimensional symmetrized version of the Rashba Hamiltonian,[33] and neglect intersubband mixing which is permissible if $W << \hbar^2/\alpha_R m^*_s$, where $W$ is the width of the transverse confining potential that defines the channel.[3,34,35] $\alpha_R$ is the spin-orbit Rashba parameter. $\delta E_c$ models the conduction band mismatch between the semiconductor and ferromagnets. In the ferromagnetic metal the energy spectrum is $E^\sigma_f(k_f) = \frac{\hbar^2}{2m^*_f}k_f^2 + \frac{1}{2}\lambda_\sigma\Delta$ where $\sigma =\uparrow, \downarrow$ indicates the spin-state of the split band and $\lambda_{\uparrow,\downarrow} = \pm 1$, the axis of spin quantization being the $z$-axis. In the semiconductor there is a Rashba splitting of the dispersion linear in $k$, $E^\sigma_s(k_s) = \frac{\hbar^2}{2m^*_s}k_s^2 + \lambda_\sigma \alpha_R k_s + \delta E_c$.

Since the Hamiltonian (5) is spin-diagonal we consider eigenstates of the whole F/S/F structure of the form $|\Psi_\uparrow\rangle = [\psi_\uparrow(x), 0], |\Psi_\downarrow\rangle = [0, \psi_\downarrow(x)]$. The matching conditions for the wave functions at the interfaces at $x_0 = 0$ and $x_0 = l_s$ are obtained by integrating $H|\Psi_\sigma\rangle = E|\Psi_\sigma\rangle$ from $x_0 - \epsilon$ to $x_0 + \epsilon$ in the limit $\epsilon \to 0$. This yields

$$\mu \frac{\partial}{\partial x}\psi^f_\sigma(x)|_{x=x_o} = \frac{\partial}{\partial x}\psi^s_\sigma(x)|_{x=x_o} + i\lambda_\sigma k_R \psi^s_\sigma(x_o) \quad (6)$$

$$\psi^f_\sigma(x_o) = \psi^s_\sigma(x_o) \quad (7)$$

with $\mu \equiv m^*_s/m^*_f$, and $k_R = m^*_s \alpha_R/\hbar^2$. In the ferromagnetic regions the eigenstates have the form

$$\psi^{f,\nu}_\sigma(x) = A^\nu_\sigma e^{ik^\nu_{F\sigma}x} + B^\nu_\sigma e^{-ik^\nu_{F\sigma}x} \quad (8)$$

with $\nu = L, R$ denoting the left and right ferromagnets. $k^\nu_{F\sigma}$ is the Fermi wave vector for the band with spin state $\sigma$ in the ferromagnet $\nu$. In the semiconductor

$$\psi^s_\uparrow(x) = C_\uparrow e^{ik^s_{F\uparrow}x} + D_\uparrow e^{-ik^s_{F\downarrow}x} \quad (9)$$

$$\psi^s_\downarrow(x) = C_\downarrow e^{ik^s_{F\downarrow}x} + D_\downarrow e^{-ik^s_{F\uparrow}x} \quad (10)$$

$k^s_{F\sigma}$ being the Fermi wave vector in the semiconductor for the spin-orbit-split band with spin $\sigma$. For the parallel (P) magnetic configuration, i.e., when the orientations of the magnetic moments of the left (L) and right (R) ferromagnets are parallel $[\vec{m}_L = \vec{m}_R = (0, 0, 1)]$, the coefficients $A^\nu_\sigma, B^\nu_\sigma, C_\sigma$, and $D_\sigma$ are determined by applying the boundary conditions (6) and (7). The probability of an incoming electron from the left ferromagnet at the Fermi energy $E_F$ in spin state $\sigma$ being transmitted to the right ferromagnet is determined by $T^P_\sigma = (k^R_{F\sigma}/k^L_{F\sigma})|A^R_\sigma|^2/|A^L_\sigma|^2$, with $B^R_\sigma = 0$. Explicitly,

$$T_\sigma^P = \frac{4\mu^2 k_{F\sigma}^L k_{F\sigma}^R (k_{F\uparrow}^s + k_{F\downarrow}^s)^2}{\kappa_{\sigma_+}^2 + \kappa_{\sigma_-}^2 - 2\kappa_{\sigma_+}\kappa_{\sigma_-}\cos[(k_{F\uparrow}^s + k_{F\downarrow}^s)l_s]} \quad (11)$$

with the definitions $\kappa_{\sigma_\pm} \equiv (K_s \pm \mu k_{F\sigma}^L)(K_s \pm \mu k_{F\sigma}^R)$, where by energy conservation, $K_s \equiv k_{F\sigma}^s + \lambda_\sigma k_R = \sqrt{k_R^2 + \mu(k_{F\sigma}^L)^2 - \frac{2m_s^*}{\hbar^2}(\delta E_c - \frac{1}{2}\lambda_\sigma \Delta)}$. For the antiparallel magnetization $(AP)$, i.e., $\vec{m}_R = -\vec{m}_L = (0,0,-1)$, the transmission probabilities $T_\uparrow^{AP}$ and $T_\downarrow^{AP}$ are given by Eq. (11) with the replacements $k_{F\uparrow}^R \to k_{F\downarrow}^L$ and $k_{F\downarrow}^R \to k_{F\uparrow}^L$, respectively. Notice that $T_\uparrow^{AP} = T_\downarrow^{AP}$ by symmetry as no external magnetic fields are considered.

The spin-conductances are then calculated within the Landauer formalism of ballistic transport,[28]

$$G^{P/AP} = \frac{e^2}{h}\sum_\sigma T_\sigma^{P/AP}. \quad (12)$$

The heavy solid line in Fig. 1(b) shows $\eta = \Delta G/2G_{av}$, the normalized change in conductance between the parallel and anti-parallel configurations of the magnetic moments of the ferromagnetic electrodes, plotted against $k_R/k_o$ ($k_o \equiv 1\times 10^5\ cm^{-1}$) for a Fe/InAs/Fe F/S/F structure with $l_s = 0.1\mu$m. Notice that a change in the sign of $\Delta G/2G_{av}$ occurs at $k_R = 1.4k_o$ as the Rashba spin-orbit coupling strength is varied, which can be accomplished experimentally by means of gating[32]. The important conclusion here is that the accepted (semi-classical) spintronic interpretation of conductance measurements must be reconsidered in the coherent quantum regime: At this value of $k_R$, the conductance change between P and AP magnetic configurations is zero ($\eta = \Delta G/2G_{av} = 0$); thus the standard semi-classical expectation is that there should be *no* spin-injection (compare Eq. (2) and (4)) at this value of $k_R$. However there is an imbalance of the spin transmission probabilities since $T_\uparrow^P \neq T_\downarrow^P$ although $T_\uparrow^{AP} = T_\downarrow^{AP}$. Thus we find that *in the coherent quantum regime* finite spin-injection can occur for the parallel configuration of ferromagnetic electrodes despite $\Delta G/2G_{av}$ being zero, contrary to semi-classical intuition.

We note in passing that $T_\uparrow^P \neq T_\downarrow^P$ for $k_R = 0$ which implies that (as in the semi-classical regime[26]) spin-injection can also occur in the absence of the Rashba coupling. The quantum interference that tunes $\eta$ (and the spin transmission probabilities) is exhibited more clearly in the inset where oscillatory behavior of $\eta$ with $l_s$ is seen.

The results shown in Fig.2 for a larger channel length ($l_s = 1\mu$m) are even more interesting. For $k_R = 2.4k_o$ and $k_R = 3.8k_o$ we again find $\eta = 0$ (Fig. 2(a)) with finite spin injection (Fig. 2(b)). But now a *maximum* of $\eta$ occurs at $k_R = 3.2k_o$ where $T_\uparrow^P = T_\downarrow^P$ and $T_\uparrow^{AP} = T_\downarrow^{AP}$. This means that at $k_R = 3.2k_o$ there is a pronounced spin valve effect, *i.e.*, a change in the conductance between the parallel and anti-parallel configurations of the magnetizations of the contacts, although *no* net spin current flows. This is an inherently *quantum* spin valve effect since it is *maximal* where the spin injection *vanishes* whereas semi-classical reasoning predicts (compare Eqs. (2) and (4)) that there should be *no* spin valve effect whenever no spin injection occurs.

In summary, we have presented calculations of ballistic electron spin transport in ferromagnetic metal /semiconductor/ferromagnetic metal structures in the coherent quantum regime. Our results demonstrate that in the coherent quantum regime the relationship between spin transport and conductance measurements (a key experimental probe of spintronic phenomena) is *qualitatively* different than in the semiclassical regime that has been studied experimentally to date: In the quantum regime a comparison of the conductances of a heterostructure with parallel and antiparallel magnetizations of magnetic contacts can no longer be regarded as an unequivocal indicator as to whether or not spin injection is taking place; it should be supplemented by other probes in studies of coherent spin injection. Moreover, we predict that coherent quantum systems should exhibit an unexpected quantum spin-valve effect that occurs even in the absence of a net spin current flowing through the device.[36] These surprising conclusions do *not* rely on the semiconductor-specific Rashba spin-orbit coupling that we include in our model Hamiltonian, but are general consequences of quantum interference.[37,38] They should apply to all-metal and all-semiconductor systems as well as to the ferromagnetic metal /semiconductor/ferromagnetic metal heterojunctions that we have discussed. Differences between the coherent quantum regime and the semi-classical regime such as those that we have described should also occur if potential barriers are present at the interfaces between the ferromagnetic and non-magnetic parts of the structure.[38] They should be taken into consideration in interpreting spin injection experiments in the quantum regime and in schemes for quantum computation that involve spin injection.

* Present address: *Centro de Ciencias de la Materia Condensada UNAM, 22800 Ensenada BC, México.* E-mail: *fmireles@ccmc.unam.mx*.

## ACKNOWLEDGMENTS

This work was supported by NSERC and by the Canadian Institute for Advanced Research.

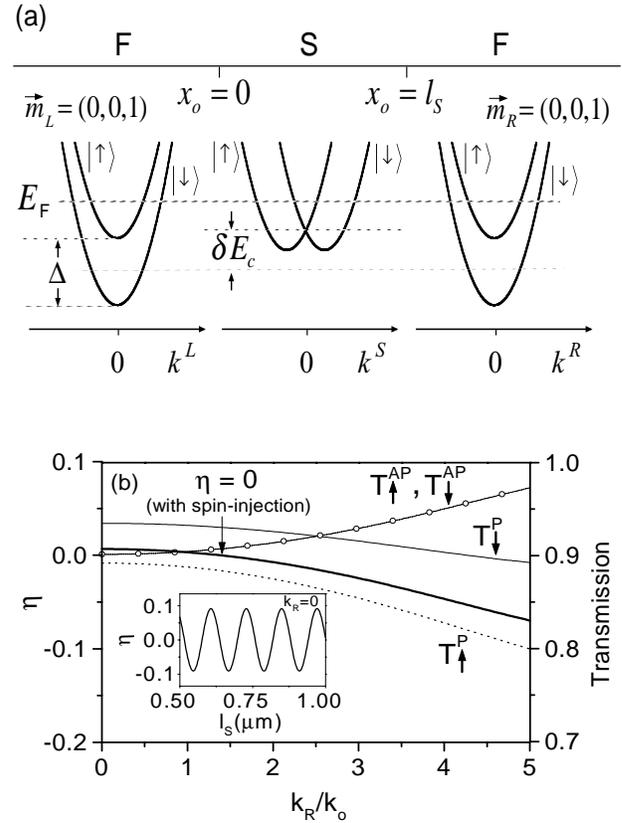

FIG. 1. (a) Schematic diagram of the split bands in a F/S/F heterojunction. (b) Transmission probability (right vertical axis) and normalized change in conductance between the parallel and anti-parallel configurations of the magnetic moments of the two ferromagnetic electrodes (left vertical axis) against $k_R/k_o$ for $l_s = 0.1\mu m$. For the ferromagnets the Fermi wave vectors are $k_{F\downarrow} = 1.05 \times 10^8\, cm^{-1}$ and $k_{F\uparrow} = 0.44 \times 10^8\, cm^{-1}$ appropriate for Fe. The effective masses were set to $m_f^* = m_e$ and $m_s^* = 0.036 m_e$ for InAs. The exchange splitting energy in the ferromagnets has been set to $\Delta = 3.46\, eV$, with a band mismatch $\delta E_c = 2.4\, eV$. For typical electron densities $n_s = 1 \times 10^{12} cm^{-2}$ the Fermi wave vector of the InAs-based semiconductor (in the absence of the Rashba coupling) is $k_F^s = \sqrt{2\pi n_s} \approx 2.5 \times 10^6 cm^{-1}$.

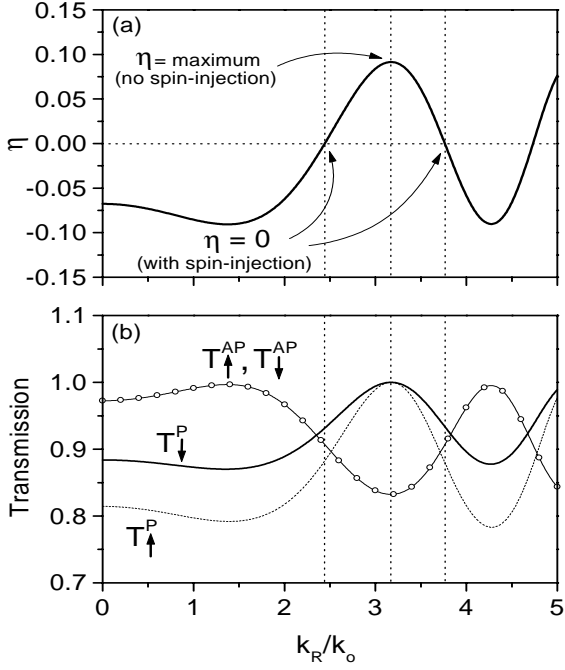

FIG. 2. (a) Quantum spin valve signal versus $k_R/ko$ for $l_s = 1\mu m$; the rest of the parameters are as in Fig. 1. The dashed vertical lines are to guide the eye. (b) Plots of transmission probabilities for both magnetization configurations.